\documentclass[]{JHEP3}

\usepackage{latexsym} 	
\usepackage{amssymb}  	
\usepackage{amsbsy}
\usepackage{epsfig}     

\newcommand{\eps}{\epsilon}
\newcommand{\vareps}{\varepsilon}

\newcommand{\om}{\omega}

\newcommand{\C}{{\cal C}}

\newcommand{\bra}{\langle}
\newcommand{\ket}{\rangle}

\newcommand{\half}{\frac{1}{2}}
\newcommand{\re}{\mbox{Re\,}}
\newcommand{\im}{\mbox{Im\,}}

\newcommand{\vecp}{{\mathbf p}}

\newcommand{\vecnul}{{\mathbf 0}}  
  
\newcommand{\pv}{{\mathbf p}}

\newcommand{\rmR}{{\rm R}}
\newcommand{\rmI}{{\rm I}}

\newcommand{\cA}{{\cal A}}

\newcommand{\cC}{{\cal C}}

\newcommand{\cM}{{\cal M}}

\newcommand{\be}{\begin{equation}}
\newcommand{\ee}{\end{equation}}
\newcommand{\bea}{\begin{eqnarray}}
\newcommand{\eea}{\end{eqnarray}}
\newcommand{\bean}{\begin{eqnarray*}}
\newcommand{\eean}{\end{eqnarray*}}
\newcommand{\nn}{\nonumber}

\newcommand{\bit}{\begin{itemize}}
\newcommand{\eit}{\end{itemize}}

\renewcommand{\theequation}{\arabic{section}.\arabic{equation}}

\title{Complex Langevin dynamics at finite chemical potential: 
 mean field analysis in the relativistic Bose gas}

\author{
 Gert Aarts \\
 Department of Physics, Swansea University, Swansea, United Kingdom \\
 Email: \email{g.aarts@swan.ac.uk}
}

\abstract{
 Stochastic quantization can potentially be used to simulate theories with 
a complex action due to a nonzero chemical potential. We study complex 
Langevin dynamics in the relativistic Bose gas analytically, using a mean 
field approximation. We concentrate on the region with a Silver Blaze 
problem and discuss convergence, stability, fixed points, and the 
severeness of the sign problem. The real distribution satisfying the 
extended Fokker-Planck equation is constructed and its nonlocal form is 
explained. Finally, we compare the mean field results in finite volume 
with the numerical data presented in Ref.~\cite{Aarts:2008wh}.
 }


\preprint{arXiv:0902.4686 [hep-lat]}

\begin{document}



\section{Introduction}
\label{sec:Introduction}
\setcounter{equation}{0}

Theories with a complex action are not easy to solve numerically, since 
approaches based on importance sampling break down. This is commonly 
referred to as the sign problem. An important theory in this class is QCD 
at finite baryon chemical potential, with a complex fermion determinant 
satisfying $[\det M(\mu)]^*=\det M(-\mu)$.\footnote{In case of a complex 
chemical potential, this relation becomes $[\det M(\mu)]^*=\det 
M(-\mu^*)$.}
  Several methods have been devised to circumvent the sign problem in QCD, 
mostly at small chemical potential and in the vicinity of the crossover 
between the confined and the deconfined phase 
\cite{Fodor:2001au,Fodor:2001pe,Fodor:2002km,Fodor:2004nz,Allton:2002zi,Allton:2003vx,Allton:2005gk,Gavai:2003mf,de 
Forcrand:2002ci,de 
Forcrand:2003hx,deForcrand:2006pv,D'Elia:2002gd,Fodor:2007vv}. For a 
detailed lattice QCD study of the sign problem at small chemical 
potential, see Ref.\ \cite{Ejiri:2008xt}. Considerable insight in the QCD 
sign problem has also been obtained with Random Matrix Theory 
\cite{Osborn:2004rf,Akemann:2004dr,Osborn:2005ss,Splittorff:2006fu,Han:2008xj,Bloch:2008cf}. 
In some theories the sign problem can be eliminated altogether, using a 
reformulation in terms of different degrees of freedom 
\cite{Chandrasekharan:1999cm,Endres:2006xu,Chandrasekharan:2008gp}.

Since stochastic quantization \cite{Parisi:1980ys} does not rely on 
importance sampling, it can potentially be applied to theories with a 
complex action using complex Langevin dynamics 
\cite{Parisi:1984cs,Klauder:1985a}. Studies in the 80's, however, have 
given mixed results, see e.g.\ Refs.\ \cite{Karsch:1985cb,Ambjorn:1986fz}. 
For an extensive review and more references, see Ref.\ 
\cite{Damgaard:1987rr}. Recently the approach was reconsidered as a method 
to solve nonequilibrium quantum fields dynamics in Minkowski spacetime 
\cite{Berges:2005yt,Berges:2006xc,Berges:2007nr}. It was shown that 
instabilities, which plagued earlier studies, can be controlled by using 
small enough Langevin stepsizes. Moreover, insight in the convergence 
properties of the method can be obtained from features of classical flow 
diagrams. Other recent applications include PT symmetric theories 
\cite{Bernard:2001wh} and unbounded actions \cite{Pehlevan:2007eq}. In 
Ref.\ \cite{Aarts:2008rr} we applied stochastic quantization to various 
theories with a nonzero chemical potential.  In particular, we considered 
QCD with static quarks, in which the fermion determinant is approximated 
but the full gauge dynamics is preserved. First results on a $4^4$ lattice 
are encouraging. The required extension from SU(3) to SL(3,$\mathbb{C}$) 
is discussed in detail.

The sign problem in QCD at finite chemical potential does not arise 
because of the anticommuting nature of the quark fields. Also 
in bosonic theories with a nonzero chemical potential and an 
action that behaves under complex conjugation as $S^*(\mu)=S(-\mu)$, the 
sign problem appears. In 
Ref.\ \cite{Aarts:2008wh} we considered the relativistic Bose gas (a self 
interacting complex scalar field) in four dimensions in the presence of a 
chemical potential as one of the simplest examples of a relativistic 
field theory with a severe sign problem. Like QCD, this theory has a 
Silver Blaze problem \cite{Cohen:2003kd}: at strictly zero temperature and 
small chemical potential, bulk physical observables are independent of the 
chemical potential, even though it enters explicitly in the microscopic 
dynamics. At larger chemical potential, the system enters a Bose condensed 
phase. The $\mu$ independence below onset and the formation of a state 
with nonzero density above onset is similar to what is expected to occur 
in QCD at zero temperature. It was demonstrated in Ref.\ 
\cite{Aarts:2008wh} that complex Langevin dynamics reproduces the expected 
physics, on lattices of size $N^4$, with $N=4,6,8,10$. The sign problem 
was shown to be severe. However, no obstacles related to the sign problem, 
the Silver Blaze problem, or in taking the thermodynamic limit were 
encountered.

In this paper we complement the numerical study of Ref.\ 
\cite{Aarts:2008wh} with a detailed analytical study in the mean field 
approximation. We concentrate on the region with the Silver Blaze problem. 
The paper is organized as follows. In Sec.\ \ref{sec:model} we remind the 
reader of the model and the corresponding complex Langevin equations. In 
order to prepare for the mean field analysis, we first discuss the case 
without interactions. In Sec.\ \ref{sec:free} we summarize the exact 
results in the free field limit, using standard field theory. Subsequently 
the free Langevin equations are solved analytically, both for continuous 
and discretized dynamics. We discuss convergence and stability properties. 
In Sec.\ \ref{sec:fokker} the stationary solution of the Fokker-Planck 
equation is given, again ignoring interactions, and shown to be in 
agreement with the solution of the Langevin equations in the limit of 
large Langevin time.\footnote{Recently, an interesting approach to study 
stationary solutions of complex Langevin dynamics was presented in Ref.\ 
\cite{Guralnik:2009pk}.} The criterium for convergence is derived from 
this distribution as well. We find that the real probability distribution 
is highly nonlocal and explain why. In Sec.\ \ref{sec:mean} interactions 
are included and the analysis is extended to the mean field approximation. 
We derive fixed points of the mean field Langevin equations at finite 
Langevin stepsize. Finally the mean field predictions in finite volume are 
compared with the nonperturbative results obtained by complex Langevin 
simulations \cite{Aarts:2008wh}. The appendix contains a short remark 
about lattice dispersion relations.

\section{Relativistic Bose gas and Langevin dynamics}
\label{sec:model}
\setcounter{equation}{0}

We consider a self-interacting complex scalar field in the presence of a 
chemical potential $\mu$, with the continuum action
 \be
 S = \int d^4x\,\left[ |\partial_\nu\phi|^2
 + (m^2-\mu^2)|\phi|^2 
 + \mu\left(\phi^*\partial_4\phi - \partial_4\phi^* \phi \right)
 + \lambda|\phi|^4 \right].
\ee
 The euclidean action is complex and satisfies $S^*(\mu) = S(-\mu)$. We take $m^2>0$, so that at 
 vanishing  and small $\mu$ the theory is in its symmetric phase.

 We study this theory on the lattice, with the action
\be
 S = \sum_x \bigg[ \left(2d+m^2\right) \phi_x^*\phi_x 
 + \lambda\left( \phi_x^*\phi_x\right)^2
- \sum_{\nu=1}^4\left(  \phi_x^* e^{-\mu\delta_{\nu,4}} \phi_{x+\hat\nu} 
+ \phi_{x+\hat\nu}^* e^{\mu\delta_{\nu,4}} \phi_x \right)
\bigg].
\ee
 As always, chemical potential is introduced as an imaginary constant 
vector 
potential in the temporal direction \cite{Hasenfratz:1983ba}.
 The number of euclidean dimensions is $d=4$, the lattice spacing $a_{\rm 
lat}\equiv 1$, and the lattice four-volume is $\Omega=N_s^3N_\tau$, where 
$N_s$ ($N_\tau$) are the number of sites in a spatial (temporal)  
direction. We use periodic boundary conditions.

 In order to formulate the complex Langevin equations for this theory, the 
complex field is first written in terms of two real fields $\phi_a$ 
($a=1,2$) as $\phi=\frac{1}{\sqrt{2}}(\phi_1+i\phi_2)$. The lattice action 
then reads
 \bea
 S 
 = \sum_x\bigg[ \half\left(
 2d+m^2\right) \phi_{a,x}^2
 + \frac{\lambda}{4}\left(\phi_{a,x}^2\right)^2
 - \sum_{i=1}^3 \phi_{a, x}\phi_{a, x+\hat i}
&& \nn \\
 -\cosh\mu\,  \phi_{a, x}\phi_{a, x+\hat 4}
 +i\sinh\mu\, \vareps_{ab}\phi_{a, x}\phi_{b, x+\hat 4}
\bigg]. &&
\label{eq:S}
\eea
 We use the antisymmetric tensor $\vareps_{ab}$, with 
$\vareps_{12}=-\vareps_{21}=1$, $\vareps_{11}=\vareps_{22}=0$, 
and summation over repeated indices is implied throughout. 

Since the Boltzmann weight $e^{-S}$ in the partition function,
\be
\label{eq:Z1}
Z = \int D\phi_1 D\phi_2\, e^{-S},
\ee
 is complex, the theory has a sign problem and one 
cannot rely on importance sampling. Writing the weight as $|e^{-S}| e^{i\varphi} = e^{-S_R}e^{-iS_I}$, one may consider the phase quenched theory
\be
\label{eq:Z2}
Z_{\rm pq} = \int D\phi_1 D\phi_2\, |e^{-S}|
= \int D\phi_1 D\phi_2\, e^{-S_{R}},
\ee
 where  $S_{R}$ is the real part of the action in Eq.\ (\ref{eq:S}), i.e.\ the 
term proportional to $\sinh\mu$ is dropped. By analysing the average phase
factor in the phase quenched theory, $\bra e^{i\varphi}\ket_{\rm pq}$, it 
was shown in Ref.\ \cite{Aarts:2008wh} 
that this theory has a severe sign problem: at nonzero chemical 
potential the average phase factor goes to zero exponentially fast in the 
thermodynamic limit.

We use stochastic quantization. The Langevin 
equations for the fields $\phi_a$ read
\be
\label{eq:CL1}
\frac{\partial }{\partial\theta}\phi_{a,x}(\theta)  =
 -\frac{\delta S[\phi]}{\delta\phi_{a,x}(\theta)} + \eta_{a,x}(\theta),
\ee
where $\theta$ is the Langevin time. The noise $\eta$ is Gaussian and normalized as
\be
\bra \eta_{a,x}(\theta)\ket = 0,
\;\;\;\;\;\;\;\;\;
\bra \eta_{a,x}(\theta)\eta_{b,x'}(\theta')\ket =
2\delta_{ab}\delta_{xx'}\delta(\theta-\theta').
\ee
 Since the force in Eq.\ (\ref{eq:CL1}) is complex, the fields are complexified as
\be
\label{eqphiRI}
 \phi_a\to \phi_a^\rmR +i\phi_a^\rmI \;\;\;\;\;\;\;\;(a=1,2).
\ee 
 The complex Langevin equations we consider in this paper then read
\bea
 \label{eqphiR}
\frac{\partial}{\partial\theta}\phi_{a,x}^\rmR(\theta)  &=& 
K_{a,x}^\rmR(\theta) + \eta_{a,x}(\theta),
\\
 \label{eqphiI}
\frac{\partial}{\partial\theta}\phi_{a,x}^\rmI(\theta) &=& 
K_{a,x}^\rmI(\theta).
\eea
 The noise is chosen to be real.
The drift terms are defined as
 \bea
 K_{a,x}^\rmR &=&  -\re \frac{\delta 
S}{\delta\phi_{a,x}}\Big|_{\phi_a\to
\phi_a^\rmR+i\phi_a^\rmI},
\\
 K_{a,x}^\rmI &=& -\im \frac{\delta 
S}{\delta\phi_{a,x}}\Big|_{\phi_a\to
\phi_a^\rmR+i\phi_a^\rmI},
\eea
and read explicitly
\bea
\label{eq:KRf}
 K_{a,x}^\rmR &=&
 -\left[ 2d + m^2 +\lambda\left( \phi_{b,x}^{\rmR\,2} -
 \phi_{b,x}^{\rmI\,2} \right) \right] \phi_{a,x}^\rmR
  +2\lambda \phi_{b,x}^\rmR\phi_{b,x}^\rmI \phi_{a,x}^\rmI
 + \sum_i\left(\phi_{a,x+\hat i}^\rmR + \phi_{a,x-\hat i}^\rmR \right)
 \nn \\ &&
 + \cosh \mu \left(\phi_{a,x+\hat 4}^\rmR + \phi_{a,x-\hat 4}^\rmR \right)
  + \sinh\mu \,\,\vareps_{ab}
        \left(\phi_{b,x+\hat 4}^\rmI - \phi_{b,x-\hat 4}^\rmI \right),
 \\
\label{eq:KIf}
K_{a,x}^\rmI &=&
-\left[ 2d + m^2  +\lambda\left(
\phi_{b,x}^{\rmR\,2} - \phi_{b,x}^{\rmI\,2} \right) \right]\phi_{a,x}^\rmI
  -2\lambda \phi_{b,x}^\rmR\phi_{b,x}^\rmI \phi_{a,x}^\rmR
 + \sum_i\left(\phi_{a,x+\hat i}^\rmI + \phi_{a,x-\hat i}^\rmI \right)
 \nn \\ &&
 + \cosh \mu \left(\phi_{a,x+\hat 4}^\rmI + \phi_{a,x-\hat 4}^\rmI \right)
  - \sinh\mu \,\,\vareps_{ab}
        \left(\phi_{b,x+\hat 4}^\rmR - \phi_{b,x-\hat 4}^\rmR \right).
\eea
 Observables are written in terms of the complexified fields (\ref{eqphiRI}) as well. We consider the square of the field modulus, 
 \be
\label{eqvar}
|\phi|^2 = \half\phi_a^2\to
\half\left( {\phi_a^\rmR}^2 - {\phi_a^\rmI}^2 \right)
+ i \phi_a^\rmR\phi_a^\rmI,
\ee
and the density $\bra n\ket= (1/\Omega)\partial\ln Z/\partial\mu$, given 
by $n = (1/\Omega)\sum_x n_x$, with
\bea
\nn
  n_x &=& \left( \delta_{ab} \sinh\mu - i \vareps_{ab}\cosh\mu \right) 
\phi_{a,x}\phi_{b,x+\hat 4}
 \\
 &\to&
\left( \delta_{ab}\sinh\mu  - i\vareps_{ab} \cosh\mu \right)
\left(
\phi_{a,x}^\rmR\phi_{b,x+\hat 4}^\rmR - \phi_{a,x}^\rmI\phi_{b,x+\hat 4}^\rmI 
+ i\left[
\phi_{a,x}^\rmR\phi_{b,x+\hat 4}^\rmI + \phi_{a,x}^\rmI\phi_{b,x+\hat 4}^\rmR 
\right]\right).
\nn \\
&& \label{eq:dens}
\eea
 After complexification all observables have a real and 
imaginary part.

 Employing that the noise is random, we observe that the Langevin 
equations have the following symmetry,
\be
\phi^\rmR_{1,x}\to -\phi^\rmR_{1,x},   \;\;\;\;\;\;
\phi^\rmR_{2,x}\to  \phi^\rmR_{2,x},   \;\;\;\;\;\;
\phi^\rmI_{1,x}\to \phi^\rmI_{1,x},    \;\;\;\;\;\; 
\phi^\rmI_{2,x}\to -\phi^\rmI_{2,x},  
\label{eq:sym}
\ee
for all $x$ and similar with 1 and 2 interchanged.
Under this transformation, the drift terms change as
\be
K^\rmR_{1,x} \to -K^\rmR_{1,x},          \;\;\;\;\;\;
K^\rmR_{2,x} \to K^\rmR_{2,x},		\;\;\;\;\;\;
K^\rmI_{1,x} \to K^\rmI_{1,x}, 		\;\;\;\;\;\;
K^\rmI_{2,x} \to -K^\rmI_{2,x}.
\ee
Correlation functions odd under this transformation should 
vanish after noise averaging, which implies that
\be
 \bra\phi^\rmR_{1,x}\phi^\rmR_{2,y}\ket = 
 \bra\phi^\rmI_{1,x}\phi^\rmI_{2,y}\ket =
 \bra\phi^\rmR_{1,x}\phi^\rmI_{1,y}\ket =
 \bra\phi^\rmR_{2,x}\phi^\rmI_{2,y}\ket = 0.
 \ee
The nonzero combinations are
 \be
 \label{eq:struc}
 \bra\phi^\rmR_{a,x}\phi^\rmR_{b,y}\ket\sim\delta_{ab}, 
 \;\;\;\;\;\;\;\;\; 
 \bra\phi^\rmI_{a,x}\phi^\rmI_{b,y}\ket\sim\delta_{ab},  
 \;\;\;\;\;\;\;\;\;
 \bra\phi^\rmR_{a,x}\phi^\rmI_{b,y}\ket\sim\vareps_{ab}.
  \ee
 Applying this to the expectation values of the observables in Eqs.\ 
(\ref{eqvar}, \ref{eq:dens}), we find that they are purely real. This is 
indeed what was observed numerically in Ref.\ \cite{Aarts:2008wh}.

\section{Ignoring interactions}
\label{sec:free}
\setcounter{equation}{0}

In order to set the stage for the mean field analysis, we first solve
the Langevin dynamics without interactions ($\lambda=0$), 
allowing for a detailed understanding of convergence and 
stability properties in the Silver Blaze regime.

\subsection{Standard results}

We start by summarizing the results obtained in the standard field theory 
approach (see e.g. Ref.\ \cite{Kapusta}). After going to momentum space, 
according to
\be
\phi_{a,x} = \sum_p e^{ipx}\phi_{a,p},
\ee
 where $p_i=2\pi n_i/N_s$, with $-N_s/2<n_i\leq N_s/2$, and $p_4=2\pi 
n_4/N_\tau$, with $-N_\tau/2<n_4\leq N_\tau/2$, the action (\ref{eq:S}) 
reads
 \be
 \label{eq:S0}
S = \sum_p  \half\phi_{a,-p}\left(\delta_{ab}A_p 
-\vareps_{ab}B_p\right)\phi_{b,p}
= \sum_p\half \phi_{a,-p} M_{ab,p} \phi_{b,p},
\ee
where
\be
 M_p = \left(
\begin{array}{cc}
A_p & -B_p \\
B_p & A_p 
\end{array}
\right),
\ee
and\footnote{In the formal continuum limit
$A_p\to m^2-\mu^2+p_4^2+\pv^2$, $B_p \to 2\mu p_4$.}
 \be
A_p= m^2 + 4\sum_{i=1}^3\sin^2\frac{p_i}{2}  + 2 \left( 1- 
\cosh \mu \cos p_4\right),
\;\;\;\;\;\;
B_p = 2 \sinh\mu \sin p_4.
\label{eq:AB}
\ee
Note that $A_{-p}=A_p$, $B_{-p}=-B_p$, and that $M_p$ is nonhermitian. 
The phase quenched theory is obtained by taking $B_p=0$. 
Up to an irrelevant constant, the logarithm of the partition function is
\be
\ln Z = -\half\sum_p\ln\det M_p = -\half\sum_p\ln(A_p^2+B_p^2).
\ee  
The observables we are interested in are given by 
\be
\label{eq:phi2}
\bra|\phi|^2\ket = -\frac{1}{\Omega}\frac{\partial \ln Z}{\partial 
m^2} = \frac{1}{\Omega}\sum_p\frac{A_p}{A_p^2+B_p^2},
\ee
and
\be
\label{eq:dens2}
\bra n\ket = \frac{1}{\Omega}\frac{\partial \ln Z}{\partial \mu}
= 
 -\frac{1}{\Omega}\sum_p \frac{A_pA_p'+B_pB_p'}{A_p^2+B_p^2},
\ee
where $A'=\partial A/\partial\mu=-2\sinh\mu\cos p_4$, $B'=\partial 
B/\partial\mu=2\cosh\mu\sin p_4$.
 
As always, the severeness of the sign problem is estimated by the average 
phase factor in the phase quenched theory, given 
by the ratio of the partition functions of the full and phase quenched 
theories (\ref{eq:Z1}, \ref{eq:Z2}),
 \be
 \bra e^{i\varphi}\ket_{\rm pq} = \frac{Z}{Z_{\rm pq}} = 
 e^{-\Omega\Delta f},
\ee
 where $\Delta f$, the difference between the corresponding free 
energy densities, is given by
\be
 \label{eq:deltaf}
 \Delta f = -\frac{1}{\Omega}\ln \frac{Z}{Z_{\rm pq}} = 
\frac{1}{2\Omega} \sum_p \ln\frac{A_p^2+B_p^2}{A_p^2}.
\ee
Note that this can be easily generalized to arbitrary powers of the phase 
factor in theories with nonvanishing phase factors.\footnote{ 
Define the partition function 
 $Z_\ell = \int D\phi_1D\phi_2\, |e^{-S}| e^{i\ell\varphi}$. 
 Then
$\bra e^{in\varphi}\ket_{\ell} = Z_{n+\ell}/Z_\ell =
 \exp(-\Omega\Delta f_{n+\ell,\ell})$,
with
\[
 \Delta f_{n+\ell,\ell} = -\frac{1}{\Omega}\ln \frac{Z_{n+\ell}}{Z_\ell} =
\frac{1}{2\Omega} \sum_p 
\ln\frac{A_p^2+(n+\ell)^2B_p^2}{A_p^2+\ell^2B_p^2}.
\]
}

Finally, since the eigenvalues of $M_p$ in the action (\ref{eq:S0}) are 
$A_p\pm iB_p$, the theory 
without interactions exists provided that $A_p>0$. This yields the 
standard stability criterium for a free Bose gas at finite chemical 
potential,
 \be
  A_p > 0  \;\;\;\; \Leftrightarrow \;\;\;\; 4\sinh^2\frac{\mu}{2} < m^2,
 \ee
 corresponding to $\mu^2<m^2$ in the formal continuum limit.
 We restrict the analysis below therefore to the case that $A_p>0$; 
 this is the Silver Blaze region.

\subsection{Continuous Langevin dynamics}

We now solve the complex Langevin equations to compare the outcome with 
the results given above. The Langevin equations (\ref{eqphiR}, \ref{eqphiI}) read in momentum space
 \bea
\frac{\partial}{\partial\theta}\phi_{a,p}^\rmR(\theta)  &=&
K_{a,p}^\rmR(\theta) + \eta_{a,p}(\theta), 
\\
\frac{\partial}{\partial\theta}\phi_{a,p}^\rmI(\theta) &=& 
K_{a,p}^\rmI(\theta),
\eea
where the noise is normalized as
\be
\bra \eta_{a,-p}(\theta)\eta_{b,p'}(\theta') \ket = 
2\delta_{ab}\delta_{pp'}\delta(\theta-\theta').
\label{eq:noise}
\ee	
Ignoring interactions, we find for the drift terms
\bea
\label{eqKR}
K_{a,p}^\rmR &=& -A_p\phi_{a,p}^\rmR +iB_p\vareps_{ab}\phi_{b,p}^\rmI,
\\
\label{eqKI}
K_{a,p}^\rmI &=& -A_p\phi_{a,p}^\rmI  - iB_p\vareps_{ab}\phi_{b,p}^\rmR,
\eea
where $A_p$ and $B_p$ are defined in Eq.\ (\ref{eq:AB}) above. 
In terms of
 \be
 \label{eq:M}
 {\cal M}_p = \left(
 \begin{array}{cccc}
 A_p & 0 &0 & -iB_p \\
 0 & A_p & iB_p & 0 \\
 0 & iB_p & A_p & 0 \\
 -iB_p & 0 & 0 & A_p
 \end{array}
\right),
\;\;\;\;
\Phi_p=\left(
 \begin{array}{c}
 \phi_{1,p}^\rmR \\ 
 \phi_{1,p}^\rmI \\ 
 \phi_{2,p}^\rmR \\ 
 \phi_{2,p}^\rmI
 \end{array}
\right), 
\;\;\;\;
 \Xi_p = \left(
 \begin{array}{c}
 \eta_{1,p} \\ 0 \\ \eta_{2,p} \\ 0
 \end{array}
\right),
 \ee
the Langevin dynamics is written as
 \be
 \frac{\partial}{\partial\theta}\Phi_p = -{\cal M}_p\Phi_p + \Xi_p.
 \ee
 The matrix ${\cal M}$ can be diagonalized by an orthogonal 
transformation and has doubly degenerate eigenvalues $\lambda_p=A_p\pm 
iB_p$. The solution of the Langevin equations is
\bea
\phi_{a,p}^{\rmR}(\theta) &=& e^{-A_p\theta}\left[ 
\cos(B_p\theta)\phi_{a,p}^\rmR(0) + 
i\sin(B_p\theta)\vareps_{ab}\phi_{b,p}^\rmI(0)\right] 
\nn\\
&& 
+ \int_0^\theta ds\, e^{-A_p(\theta-s)}\cos[B_p(\theta-s)]\eta_{a,p}(s),
\\
\phi_{a,p}^{\rmI}(\theta) &=& e^{-A_p\theta}\left[ 
\cos(B_p\theta)\phi_{a,p}^\rmI(0) -
i\sin(B_p\theta)\vareps_{ab}\phi_{b,p}^\rmR(0)\right]
\nn\\
&& 
-i \int_0^\theta ds\, 
e^{-A_p(\theta-s)}\sin[B_p(\theta-s)]\vareps_{ab}\eta_{b,p}(s),
\eea
where $\phi_{a,p}^{\rmR,\rmI}(0)$ denote the initial conditions.

We are now in a position to discuss the convergence properties of the 
Langevin process in the limit of large Langevin time. First we note that 
there is independence of initial conditions provided that $A_p>0$, i.e.\ 
in the region of interest here. Taking $\phi^{\rmR,\rmI}_{a,p}(0)=0$, we 
find for the two-point functions, after using Eq.~(\ref{eq:noise}) and 
performing the Langevin time integrals,
 \bea
\nn
\bra\phi^\rmR_{a,-p}(\theta)\phi^\rmR_{b,p'}(\theta)\ket 
&=&
 \frac{1}{2A_p}\frac{\delta_{ab}\delta_{pp'}}{A_p^2+B_p^2} 
\bigg( 2A_p^2+B_p^2 
\\ \nn  &&  
-e^{-2A_p\theta}\left[ A_p^2+B_p^2 +  
A_p^2\cos(2B_p\theta) 
- A_pB_p\sin(2B_p\theta)\right]\bigg),
\\
\nn
 \bra\phi^\rmI_{a,-p}(\theta)\phi^\rmI_{b,p'}(\theta)\ket 
 &=&
\frac{1}{2A_p}\frac{\delta_{ab}\delta_{pp'}}{A_p^2+B_p^2}
\bigg( B^2 
 \\ \nn &&
  - e^{-2A_p\theta}\left[ A_p^2+B_p^2  
- A_p^2\cos(2B_p\theta) 
+ A_pB_p\sin(2B_p\theta)\right]\bigg),
\\
\bra\phi^\rmR_{a,-p}(\theta)\phi^\rmI_{b,p'}(\theta)\ket
 &=&
 \frac{i}{2}\frac{\vareps_{ab}\delta_{pp'}}{A_p^2+B_p^2}
\bigg( B_p
- e^{-2A_p\theta}\left[
B_p\cos(2B_p\theta) + A_p\sin(2B_p\theta)\right]\bigg).
\;\;
\eea
 Most of the terms vanish in the limit that  
$\theta\to \infty$, again provided that $A_p>0$. The surviving terms are
\bea
\nn
\lim_{\theta\to\infty} 
\bra\phi^\rmR_{a,-p}(\theta)\phi^\rmR_{b,p'}(\theta)\ket &\equiv&
\bra\phi^\rmR_{a,-p}\phi^\rmR_{b,p'}\ket
 = 
 \delta_{ab} \delta_{pp'} \frac{1}{2A_p}\frac{2A_p^2+B_p^2}{A_p^2+B_p^2},\\
\nn 
\lim_{\theta\to\infty} 
\bra\phi^\rmI_{a,-p}(\theta)\phi^\rmI_{b,p'}(\theta)\ket &\equiv&
\bra\phi^\rmI_{a,-p}\phi^\rmI_{b,p'}\ket
=
  \delta_{ab} \delta_{pp'} \frac{1}{2A_p}\frac{B_p^2}{A_p^2+B_p^2},\\
\lim_{\theta\to\infty}  
\bra\phi^\rmR_{a,-p}(\theta)\phi^\rmI_{b,p'}(\theta)\ket &\equiv&
\bra\phi^\rmR_{a,-p}\phi^\rmI_{b,p'}\ket
=
  \vareps_{ab} \delta_{pp'} \frac{i}{2}\frac{B_p}{A_p^2+B_p^2}.
\label{eq:two}
\eea
 The structure of these two-point functions is in agreement with 
the symmetry (\ref{eq:sym}, \ref{eq:struc}) discussed above. 

For the observables we find the following.
The square of the field modulus (\ref{eqvar}) is given by
\bea
\nn
\bra |\phi|^2 \ket 
&=& \frac{1}{2\Omega}\sum_p \left\bra
 \phi^\rmR_{a,-p}\phi^\rmR_{a,p} -\phi^\rmI_{a,-p}\phi^\rmI_{a,p} + 2i 
\phi^\rmR_{a,-p}\phi^\rmI_{a,p}
\right\ket
\nn \\ 
&=&  \frac{1}{\Omega}\sum_p \frac{A_p}{A_p^2+B_p^2},
\eea
which agrees with Eq.\ (\ref{eq:phi2}).
After going to momentum space, the density (\ref{eq:dens}) reads
\bea
\bra n\ket &=&
\frac{1}{\Omega}\sum_p\left(
\delta_{ab}\sinh\mu\cos p_4 + \vareps_{ab}\cosh\mu\sin p_4  \right)
\left\bra
 \phi^\rmR_{a,-p}\phi^\rmR_{b,p} - \phi^\rmI_{a,-p}\phi^\rmI_{b,p} + 
2i\phi^\rmR_{a,-p}\phi^\rmI_{b,p}
\right\ket
\nn \\
&=&
\frac{2}{\Omega}\sum_p\left[
\sinh\mu\cos p_4 \frac{A_p}{A_p^2+B_p^2}
-\cosh\mu\sin p_4\frac{B_p}{A_p^2+B_p^2}\right],
\eea
 which agrees with Eq.\ (\ref{eq:dens2}). Note that all two-point 
functions in Eq.\ (\ref{eq:two}) contribute to this answer.

We conclude therefore that the Langevin process is independent of initial 
conditions and converges to the correct result in the limit of infinite 
Langevin time, provided that $A_p>0$, as required in the Silver Blaze 
region. Moreover, the complexification is essential, as exemplified by the observables above.

\subsection{Discretized Langevin dynamics}

We proceed by briefly considering the Langevin process after discretizing 
Langevin time as $\theta=n\eps$, where $\eps$ is the Langevin time step.
The discretized Langevin equations are 
\bea
\label{eq:phiRn}
 \phi_{a,p}^\rmR(n+1) &=& \phi_{a,p}^\rmR(n) +\eps K_{a,p}^\rmR(n) 
+\sqrt{\eps}\eta_{a,p}(n), \\
\label{eq:phiIn}
 \phi_{a,p}^\rmI(n+1) &=& \phi_{a,p}^\rmI(n) +\eps 
K_{a,p}^\rmI(n),
\eea
and the noise obeys $\bra \eta_{a,-p}(n)\eta_{b,p'}(n')\ket = 2\delta_{nn'}\delta_{ab}\delta_{pp'}$.
In the notation of Eq.\  (\ref{eq:M}), these equations are summarized 
as
\be
 \Phi_p(n+1) = \left(1-\eps {\cal M}_p\right)\Phi_p(n) + 
\sqrt{\eps}\,\Xi_p(n),
 \ee
 and solved by
\be
 \label{eq:soldis}
 \Phi_p(n) = \left(1-\eps {\cal M}_p\right)^n\Phi_p(0) +
\sqrt{\eps}\sum_{i=0}^{n-1} \left(1-\eps {\cal 
M}_p\right)^{n-1-i}\Xi_p(i),
\ee
 where $\Phi_p(0)$ is the initial condition.
 Convergence is determined by the eigenvalues of $\cM_p$. This yields the 
condition
\be
 |1-\eps\lambda_p|<1, 
\;\;\;\;\;\;\;\;
\lambda_p = A_p\pm iB_p,
\ee
resulting in the constraint
\be
\label{eq:ssc}
A_p-\frac{\eps}{2}\left(A_p^2+B_p^2\right) > 0.
\ee
 We find that the convergence criterium is modified by an explicit 
stepsize dependence. However, this restriction is not special for the 
complex Langevin process \cite{Batrouni:1985jn,Damgaard:1987rr}. 
Consider real Langevin dynamics at zero $\mu$ or in the  
phase quenched theory. In both cases $B_p=0$ and the criterium reads
\be
 0< \frac{\eps}{2} A_p <1.
\ee 
 Since $A_p$ is maximal at the edge of the Brillouin zone ($p=\pi$), this yields the modest constraint (for $\mu=0$)
\be
 \eps< \frac{2}{4d+ m^2}.
\ee
At nonzero chemical potential, this constraint  is modified to
\be
 \eps < \frac{2}{4d + m^2+2(\cosh\mu-1)},
\ee
 both in the full and the phase quenched theory. 
 In the Silver Blaze region, where $\mu$ is bounded,
 this leads to only a slightly stronger bound on $\eps$.
However, since this bound is determined by ultraviolet modes at the scale of the lattice cutoff, it is likely that 
a similar constraint holds in the high-density phase as well. In the limit that $\mu\gg 1$, exponentially small stepsizes would eventually  be required. It should be noted, however, that 
for such large chemical potentials lattice artefacts are severe.

The solution (\ref{eq:soldis}) can be used to study finite stepsize 
effects in two-point functions at infinite Langevin time. 
We come back to this below using a more elegant approach 
based on fixed points of the Langevin equations.

\section{Fokker-Planck equation}
\label{sec:fokker}
\setcounter{equation}{0}

In order to better understand the Langevin process, we now study 
properties of the associated distributions.

Consider first the Langevin process (\ref{eq:CL1}) and the distribution 
$P[\phi,\theta]$, defined via
\be
\label{eq:P1}
\bra O[\phi,\theta]\ket_\eta =
 \int D\phi\, P[\phi,\theta] O[\phi], 
\ee
where the brackets on the LHS denote noise averaging.
 The distribution satisfies the Fokker-Planck equation (in continuous Langevin time)
\be
\frac{\partial P[\phi,\theta]}{\partial\theta} = 
\sum_x \frac{\delta}{\delta\phi_{a,x}(\theta)}\left(
\frac{\delta}{\delta \phi_{a,x}(\theta)} + \frac{\delta S[\phi]}{\delta
\phi_{a,x}(\theta)}\right) P[\phi,\theta].
\ee
As always, the index $a=1,2$ is summed over.
The stationary solution, 
\be
 P[\phi]\sim e^{-S[\phi]},
\ee
always exists. However, since the action is complex, this is not the probability 
distribution for the complex Langevin process.

More relevant for the complexified process (\ref{eqphiR}, \ref{eqphiI}) we 
consider here, is the real distribution 
$\rho[\phi^\rmR,\phi^\rmI,\theta]$, defined via \cite{Parisi:1984cs}
\be
\label{eq:P2}
  \bra O[\phi^\rmR+i\phi^\rmI,\theta]\ket_\eta
=
\int D\phi^\rmR D\phi^\rmI\, \rho[\phi^\rmR,\phi^\rmI,\theta] 
O[\phi^\rmR+i\phi^\rmI].
\ee
This distribution satisfies the extended Fokker-Planck equation 
\be
\frac{\partial \rho[\phi^\rmR,\phi^\rmI,\theta]}{\partial\theta} = 
\sum_x \Bigg[ 
\frac{\delta}{\delta\phi^\rmR_{a,x}(\theta)}\left(
\frac{\delta}{\delta \phi^\rmR_{a,x}(\theta)} -K^\rmR_{a,x}(\theta)
\right) 
- \frac{\delta}{\delta\phi^\rmI_{a,x}(\theta)}
 K^\rmI_{a,x}(\theta)
\Bigg]
 \rho[\phi^\rmR,\phi^\rmI,\theta].
\label{eq:FP}
\ee
If stochastic quantization is applicable for complex actions, the two 
expectation values (\ref{eq:P1}) and (\ref{eq:P2}) should be equal \cite{Parisi:1984cs}.

We focus on the stationary solution of Eq.\ (\ref{eq:FP}) and henceforth 
drop the $\theta$ dependence. Ignoring again interactions, the stationary 
solution should satisfy
\be
\sum_p \Bigg[ 
\frac{\delta}{\delta\phi^\rmR_{a,p}}\left(
\frac{\delta}{\delta \phi^\rmR_{a,-p}} -K^\rmR_{a,p}
\right) 
- \frac{\delta}{\delta\phi^\rmI_{a,p}}
 K^\rmI_{a,p}
\Bigg]
 \rho[\phi^\rmR,\phi^\rmI] =0,
\ee
where the drift terms $K_{a,p}^{\rmR,\rmI}$ were given in Eqs.\ 
(\ref{eqKR}, \ref{eqKI}).
Explicitly, this reads
\bea
&& \sum_p \Bigg[
\frac{\delta}{\delta\phi^\rmR_{a,p}}
 \frac{\delta}{\delta \phi^\rmR_{a,-p}} 
 + \left( A_p\phi_{a,p}^\rmR -iB_p\vareps_{ab}\phi_{b,p}^\rmI \right) 
\frac{\delta}{\delta\phi^\rmR_{a,p}} 
 \nn \\  
&& \;\;\;\;\;\;\;\;
+ \left(A_p\phi_{a,p}^\rmI + 
iB_p\vareps_{ab}\phi_{b,p}^\rmR\right) \frac{\delta}{\delta\phi^\rmI_{a,p}} 
+2A_p \Bigg]
 \rho[\phi^\rmR,\phi^\rmI] =0.
\label{eq:FPmom}
\eea
Based on the structure of the equation, the solution can be written as
\be
\label{eq:rho}
\rho[\phi^\rmR,\phi^\rmI] = N\exp \left[-\sum_p\left( 
\alpha_p\phi^\rmR_{a,-p}\phi^\rmR_{a,p}
+\beta_p\phi^\rmI_{a,-p}\phi^\rmI_{a,p}
+2i\vareps_{ab}\gamma_p\phi^\rmR_{a,-p}\phi^\rmI_{b,p}
\right)\right],
\ee
where $N$ is a normalization constant.
Inserting this expression in Eq.\ (\ref{eq:FPmom}) yields the 
coefficients 
\be
\label{eq:coef}
\alpha_p = A_p, \;\;\;\;\;\;\;\;
\beta_p = \frac{A_p}{B_p^2}\left(2A_p^2+B_p^2\right), \;\;\;\;\;\;\;\;
\gamma_p = \frac{A_p^2}{B_p}.
\ee
 We have therefore found the stationary distribution corresponding to the 
complex Langevin process in the noninteracting case.\footnote{See Refs.\ 
\cite{Ambjorn:1985iw,Nakazato:1985zj} for other examples.}
 Note that since $\gamma_{-p}=-\gamma_p$, the distribution is real in 
real space, as it should be.

We now verify that this stationary solution is indeed the 
distribution corresponding to the Langevin process in the limit of 
infinite Langevin time. 
Performing the Gaussian integrals, we find the partition 
function
\be
Z = \prod_p \int d\phi^\rmR_p d\phi^\rmI_p\, \rho[\phi^\rmR,\phi^\rmI]
= {\cal N} \prod_p \frac{1}{\alpha_p\beta_p-\gamma_p^2},
\ee
where ${\cal N}$ is an irrelevant constant and
\be
\alpha_p\beta_p-\gamma_p^2 = 
\frac{A_p^2}{B_p^2}\left(A_p^2+B_p^2\right) >0.
\ee
 The two-point functions that follow from this distribution are
\bea
\nn
 \bra\phi^\rmR_{a,-p}\phi^\rmR_{a,p}\ket &=& -\frac{\partial \ln 
Z}{\partial \alpha_p} = \frac{\beta_p}{\alpha_p\beta_p-\gamma_p^2} =
\frac{1}{A_p}\frac{2A_p^2+B_p^2}{A_p^2+B_p^2},
\\
\nn
 \bra\phi^\rmI_{a,-p}\phi^\rmI_{a,p}\ket &=& -\frac{\partial \ln 
Z}{\partial \beta_p} = \frac{\alpha_p}{\alpha_p\beta_p-\gamma_p^2}
= \frac{1}{A_p}\frac{B_p^2}{A_p^2+B_p^2},
\\
2i\vareps_{ab} \bra\phi^\rmR_{a,-p}\phi^\rmI_{b,p}\ket &=& - 
\frac{\partial \ln Z}{\partial \gamma_p} 
 = \frac{-2\gamma_p}{\alpha_p\beta_p-\gamma_p^2}
= \frac{-2B_p}{A_p^2+B_p^2}.
\eea
These agree exactly with the results obtained by solving the Langevin 
equation, cf.\ Eq.\ (\ref{eq:two}).

 The theory with the probability distribution $\rho[\phi^\rmR,\phi^\rmI]$ 
exists provided that the eigenvalues of the quadratic form in Eq.\ 
(\ref{eq:rho}) are positive. We find the eigenvalues to be 
\bea
\lambda_p &=&  \half \left( 
\alpha_p+\beta_p\pm\sqrt{(\alpha_p-\beta_p)^2+4\gamma_p^2}\right) \nn
\\
&=&
\frac{A_p}{B_p^2}\sqrt{A_p^2+B_p^2}\left(\sqrt{A_p^2+B_p^2}\pm 
A_p\right).
\eea
 These are positive provided that $A_p>0$. The criterium that determines 
the convergence of the Langevin dynamics also emerges in the stationary 
solution of the extended Fokker-Planck equation, as expected.

Let us discuss some more properties of the distribution (\ref{eq:rho}). 
First we note that the distribution is highly nonlocal in real space and 
does not allow for e.g.\ a derivative expansion, due 
to the division by $B_p=2\sinh\mu\sin p_4$ in the coefficients 
(\ref{eq:coef}).  We find therefore that the complexity of the original 
local weight $e^{-S}$ has been traded for the nonlocality of the real 
probability distribution. However, this nonlocal behaviour is expected: it 
follows from the Langevin equations that the modes with $p_4=0$ are purely 
real, i.e.\ $\phi_{a,(p_4=0,\pv)}^\rmI=0$. This is enforced in the 
probability distribution by the singular behaviour as $p_4\to 0$. For the 
same reason the limit $\mu\to 0$ is singular, since there is no need to 
complexify the dynamics in this case and the distribution for the 
$\phi^\rmI$ modes should reduce to a delta function, $\delta(\phi^\rmI)$. 
These considerations fix the dependence on $B_p$.

In conclusion, we have found the stationary solution of the extended 
Fokker-Planck distribution. The real distribution is nonlocal and 
singular in the limit that $\mu, p_4\to 0$.

\section{Mean field approximation} 
\label{sec:mean} 
\setcounter{equation}{0} 

We now return to the interacting theory, with discretized Langevin time $\theta=n\eps$, and
consider the two-point functions
\bea
\nn
G_{ab,p}^{\rmR\rmR}(n) &=& 
 \bra\phi^\rmR_{a,-p}(n)\phi^\rmR_{b,p}(n)\ket,
 \\ 
\nn
 G_{ab,p}^{\rmI\rmI}(n) &=&
 \bra\phi^\rmI_{a,-p}(n)\phi^\rmI_{b,p}(n)\ket, 
 \\
G_{ab,p}^{\rmR\rmI}(n) &=&
\bra\phi^\rmR_{a,-p}(n)\phi^\rmI_{b,p}(n)\ket. 
\label{eq:tp}
\eea 
Using the Langevin equations (\ref{eq:phiRn}, \ref{eq:phiIn}), we find 
that these correlation functions evolve according to
\bea
 G_{ab,p}^{\rmR\rmR}(n+1) &=& G_{ab,p}^{\rmR\rmR}(n)  
 + \eps \bra\phi_{a,-p}^{\rmR}(n) K_{b,p}^\rmR(n) + K_{a,-p}^{\rmR}(n) 
\phi_{b,p}^\rmR(n)\ket 
\nn \\ && \nn
 +\eps^2 \bra K_{a,-p}^{\rmR}(n) K_{b,p}^\rmR(n)\ket + 
\eps\bra\eta_{a,-p}\eta_{b,p}\ket,
\\
 G_{ab,p}^{\rmI\rmI}(n+1) &=& G_{ab,p}^{\rmI\rmI}(n)  
 + \eps \bra\phi_{a,-p}^{\rmI}(n) K_{b,p}^\rmI(n) + K_{a,-p}^{\rmI}(n) 
\phi_{b,p}^\rmI(n)\ket 
\nn \\ && \nn
 +\eps^2 \bra K_{a,-p}^{\rmI}(n) K_{b,p}^\rmI(n)\ket,
\\
 G_{ab,p}^{\rmR\rmI}(n+1) &=& G_{ab,p}^{\rmR\rmI}(n)  
 + \eps \bra\phi_{a,-p}^{\rmR}(n) K_{b,p}^\rmI(n) + K_{a,-p}^{\rmR}(n) 
\phi_{b,p}^\rmI(n)\ket 
 \nn \\ && 
+\eps^2 \bra K_{a,-p}^{\rmR}(n) K_{b,p}^\rmI(n)\ket.
\eea
Here we used that $\bra \eta_{a,-p}(n)\phi_{b,p}^{\rmR,\rmI}(n)\ket =0$,
 since the fields at time $n$ do not depend on the noise at time $n$. The 
terms proportional to $\eps^2$ are finite stepsize corrections.
We then look for fixed points of the Langevin equations, and put
\be
G^{\rmR\rmR}_{ab,p}(n+1) = G^{\rmR\rmR}_{ab,p}(n),
\ee
etc. This yields the fixed point equations
\bea
 \bra\phi_{a,-p}^{\rmR} K_{b,p}^\rmR + K_{a,-p}^{\rmR} 
\phi_{b,p}^\rmR\ket 
+\eps \bra K_{a,-p}^{\rmR} K_{b,p}^\rmR\ket 
&=& -2\delta_{ab},
\nn \\
\bra\phi_{a,-p}^{\rmI} K_{b,p}^\rmI + K_{a,-p}^{\rmI} 
\phi_{b,p}^\rmI\ket  
+\eps \bra K_{a,-p}^{\rmI} K_{b,p}^\rmI\ket &=& 0,
\nn \\
\bra\phi_{a,-p}^{\rmR} K_{b,p}^\rmI + K_{a,-p}^{\rmR} 
\phi_{b,p}^\rmI\ket 
+\eps \bra K_{a,-p}^{\rmR} K_{b,p}^\rmI\ket &=& 0.
\label{eq:set}
\eea
To implement a mean field approximation and find a self-consistent set for the two-point functions 
(\ref{eq:tp}), we factorize the interaction terms. Consider for example the term $\phi_{b,x}^{\rmR}\phi_{b,x}^{\rmI}  \phi_{a,x}^\rmR$ appearing in the drift term (\ref{eq:KIf}). We write 
\be
  \phi_{b,x}^{\rmR}\phi_{b,x}^{\rmI}  \phi_{a,x}^\rmR
 \to 
 \bra\phi_{b,x}^{\rmR} \phi_{b,x}^{\rmI} \ket\phi_{a,x}^\rmR
 + \bra\phi_{b,x}^\rmR\phi_{a,x}^\rmR\ket \phi_{b,x}^\rmI 
 + \bra\phi_{b,x}^\rmI \phi_{a,x}^\rmR \ket \phi_{b,x}^\rmR.
 \ee
 Using the notation
\bea
\nn
 G^{\rmR\rmR}_{ab}(n) &=&  \bra\phi_{a,x}^\rmR(n) \phi_{b,x}^\rmR(n) 
\ket = \frac{1}{\Omega} \sum_p G_{ab,p}^{\rmR\rmR}(n), \\
\nn
 G^{\rmI\rmI}_{ab}(n) &=&  \bra\phi_{a,x}^\rmI(n) \phi_{b,x}^\rmI(n) 
\ket = \frac{1}{\Omega} \sum_p G_{ab,p}^{\rmI\rmI}(n), \\
 G^{\rmR\rmI}_{ab}(n) &=&  \bra\phi_{a,x}^\rmR(n) \phi_{b,x}^\rmI(n) 
\ket = \frac{1}{\Omega} \sum_p G_{ab,p}^{\rmR\rmI}(n),
\eea
the drift terms in the mean field approximation read
\bea
K_{a,p}^\rmR &=& 
- \left[ A_p + \lambda \left( G_{bb}^{\rmR\rmR}-G_{bb}^{\rmI\rmI} \right)\right] 
\phi_{a,p}^\rmR 
 -2\lambda  \left( G_{ab}^{\rmR\rmR}-G_{ab}^{\rmI\rmI}\right) \phi_{b,p}^\rmR 
\nn \\&& 
+ \left[ iB_p\vareps_{ab}  +2\lambda \left( G_{ab}^{\rmR\rmI}+G_{ba}^{\rmR\rmI} \right)    \right] \phi_{b,p}^\rmI
+2\lambda G_{bb}^{\rmR\rmI} \phi_{a,p}^\rmI, 
\\
K_{a,p}^\rmI &=& 
- \left[ A_p +\lambda  \left( G_{bb}^{\rmR\rmR}-G_{bb}^{\rmI\rmI} \right)   \right] \phi_{a,p}^\rmI  
-2\lambda \left( G_{ab}^{\rmR\rmR}-G_{ab}^{\rmI\rmI}\right) \phi_{b,p}^\rmI
\nn \\ &&
- \left[ iB_p\vareps_{ab} +2\lambda \left( G_{ab}^{\rmR\rmI}+G_{ba}^{\rmR\rmI} \right)     \right] \phi_{b,p}^\rmR
 -2\lambda G_{bb}^{\rmR\rmI} \phi_{a,p}^\rmR.
 \eea
These can be further simplified by noting that both $G_{ab}^{\rmR\rmR}(n)$ and $G_{ab}^{\rmI\rmI}(n)$   are proportional to $\delta_{ab}$, for all Langevin times.
We write therefore
\bea
\nn
G_{ab}^{\rmR\rmR}(n) &=& \delta_{ab}G^{\rmR\rmR}(n),
\\
\nn
G_{ab}^{\rmI\rmI}(n) &=& \delta_{ab}G^{\rmI\rmI}(n),
\\
G_{ab}^{\rmR\rmI}(n) &=& \delta_{ab} G^{\rmR\rmI}(n)+ 
\vareps_{ab} \bar G^{\rmR\rmI}(n),
\eea
 such that the drift terms reduce to
 \bea
K_{a,p}^\rmR &=& 
 -\cA_p \phi_{a,p}^\rmR 
  +i B_p\vareps_{ab} \phi_{b,p}^\rmI
 +\cC \phi_{a,p}^\rmI,
\label{eq:KRMF}
 \\
K_{a,p}^\rmI &=& 
 -\cA_p \phi_{a,p}^\rmI
  -i B_p\vareps_{ab}  \phi_{b,p}^\rmR
 -\cC \phi_{a,p}^\rmR,
\label{eq:KIMF}
  \eea
with
\be
\cA_p=
 A_p + 4\lambda \left( G^{\rmR\rmR}-G^{\rmI\rmI} \right),
\;\;\;\;\;\;\;\;
\cC = 8\lambda G^{\rmR\rmI}.
\ee
 Since $\cA_p(n)$ and $\cC(n)$ depend explicitly on the Langevin time, the time-dependent mean field  Langevin equations cannot be solved analytically.\footnote{In fact, the problem is now very similar to that of nonequilibrium field dynamics using a self-consistent mean field approximation in the equal-time formalism \cite{Aarts:2000wi}.}

We therefore look for fixed points. After substituting Eqs.\ (\ref{eq:KRMF}, \ref{eq:KIMF}) in  the fixed point equations (\ref{eq:set}) and performing some algebra, we find that at the fixed point $\C=0$ and that the  two-point functions can be decomposed as
\be
G^{\rmR\rmR}_{ab,p} = \delta_{ab}G^{\rmR\rmR}_p,
\;\;\;\;\;\;\;\;\;\;
G^{\rmI\rmI}_{ab,p} = \delta_{ab}G^{\rmI\rmI}_p,
\;\;\;\;\;\;\;\;\;\;
G^{\rmR\rmI}_{ab,p} = i\vareps_{ab}G^{\rmR\rmI}_p.
\ee
This is in agreement with the symmetry (\ref{eq:sym}, \ref{eq:struc}). The three fixed point equations (\ref{eq:set}) then become
 \bea
 \cA_pG^{\rmR\rmR}_p + B_pG^{\rmR\rmI}_p - \frac{\eps}{2}\left(
 \cA_p^2G^{\rmR\rmR}_p + B_p^2G^{\rmI\rmI}_p + 2 \cA_pB_p G^{\rmR\rmI}_p
 \right) 
 &=& 1, \nn \\
 \cA_pG^{\rmI\rmI}_p - B_pG^{\rmR\rmI}_p - \frac{\eps}{2}\left(
 \cA_p^2G^{\rmI\rmI}_p + B_p^2G^{\rmR\rmR}_p - 2 \cA_pB_p G^{\rmR\rmI}_p
\right) 
 &=& 0, 
\nn \\
 \cA_pG^{\rmR\rmI}_p - \half 
B_p\left[G^{\rmR\rmR}-G^{\rmI\rmI}_p\right]
 - \frac{\eps}{2}\left( \left[ \cA_p^2-B_p^2\right]G^{\rmR\rmI}_p
 - \cA_p B_p \left[G^{\rmR\rmR}_p-G^{\rmI\rmI}_p\right]\right) 
 &=& 0.
	\;\;\;\;\;\;\;\;\;\; 
\label{eq:fpr}
\eea
The solution is
\bea
\nn
 G^{\rmR\rmR}_p+G^{\rmI\rmI}_{p} &=& 
 \frac{1}{\cA_p-\half\eps \left(\cA_p^2+B_p^2\right)},\\
\nn
 G^{\rmR\rmR}_{p}-G^{\rmI\rmI}_{p} &=& 
 \frac{1}{\cA_p^2+B_p^2}\frac{\cA_p\left(1-\half\eps \cA_p\right) 
+ 
 \half\eps B_p^2}{\left(1-\half\eps \cA_p\right)^2 
 + \frac{1}{4}\eps^2B_p^2},\\
 G^{\rmR\rmI}_{p} &=& \frac{1}{2}\frac{B_p}{\cA_p^2+B_p^2}
 \frac{1-\eps \cA_p}{\left(1-\half\eps \cA_p\right)^2+\frac{1}{4}\eps^2 
B_p^2}.
\label{eq:ff}
\eea
 For vanishing Langevin stepsize this simplifies to
\be
 G^{\rmR\rmR}_p+G^{\rmI\rmI}_{p} = \frac{1}{\cA_p},
 \;\;\;\;\;\;\;\;
  G^{\rmR\rmR}_{p}-G^{\rmI\rmI}_{p} =  \frac{\cA_p}{\cA_p^2+B_p^2},
 \;\;\;\;\;\;\;\;
 G^{\rmR\rmI}_{p} = \frac{1}{2}\frac{B_p}{\cA_p^2+B_p^2},
\label{eq:fff}
\ee
while in the phase quenched theory ($B_p=0$) where real Langevin dynamics 
is applicable, the solution reduces to
\be
 G^{\rmR\rmR}_p = \frac{1}{\cA_p} \frac{1}{1-\half\eps \cA_p},
 \;\;\;\;\;\;\;\;\;\;
    G^{\rmI\rmI}_{p} = G^{\rmR\rmI}_{p} = 0.
\ee
 We find finite stepsize corrections linear in $\eps$, as expected 
\cite{Batrouni:1985jn}. Furthermore, we note that for large stepsize the 
denominator in the first line of Eq.\ (\ref{eq:ff}) can go negative. 
However, this occurs precisely when the stability criterium (\ref{eq:ssc}) 
is violated (after the replacement $A_p\to \cA_p$) and is therefore 
excluded.

The expressions in Eq.\ (\ref{eq:fff}) agree precisely with the solutions 
(\ref{eq:two}) obtained by solving the Langevin equations without 
interactions, after making the mean field replacement $A_p\to \cA_p$.  
This replacement corresponds to the standard mean field approximation in 
which the mass parameter receives a tadpole correction,
\be
 m^2 \to  M^2 = m^2 +4\lambda\left\bra|\phi|^2\right\ket,
\ee
or, in the notation of this section, 
\bea
 m^2 \to  M^2  &=&  m^2 + 4\lambda\left(G^{\rmR\rmR}-G^{\rmI\rmI}\right)
\nn \\ 
&=&  m^2 + \frac{4\lambda}{\Omega}\sum_p\frac{\cA_p}{\cA_p^2+B_p^2},
\eea
with
\be
 \cA_p= M^2 + 4\sum_i\sin^2\frac{p_i}{2}  + 2 \left( 1- \cosh \mu \cos 
p_4\right).
\ee
 These equations define a self-consistent gap equation for $M^2$. Given 
$m$ and $\lambda$, the gap equation can be solved numerically after 
specifying the lattice size. For example, taking $m=\lambda=1$, we find 
$M^2=1.47$ and $\bra|\phi|^2\ket = 0.119$ at $\mu=0$ on a lattice of size 
$10^4$. The critical chemical potential then follows from $\cA_0=0$ (or 
$M^2 = 4\sinh^2\frac{\mu}{2}$) and is found to be 
$\mu_c=1.15$.\footnote{When $\lambda=0$, $\mu_c$ follows from $A_0=0$ (or 
$m^2 = 4\sinh^2\frac{\mu}{2}$), yielding $\mu_c^0= 0.962$ for $m=1$.}

\FIGURE[t]{ 
\epsfig{figure=ana_full_phi2_MF.eps,height=5.3cm} 
\epsfig{figure=ana_phq_phi2_MF.eps,height=5.3cm}
 \caption{
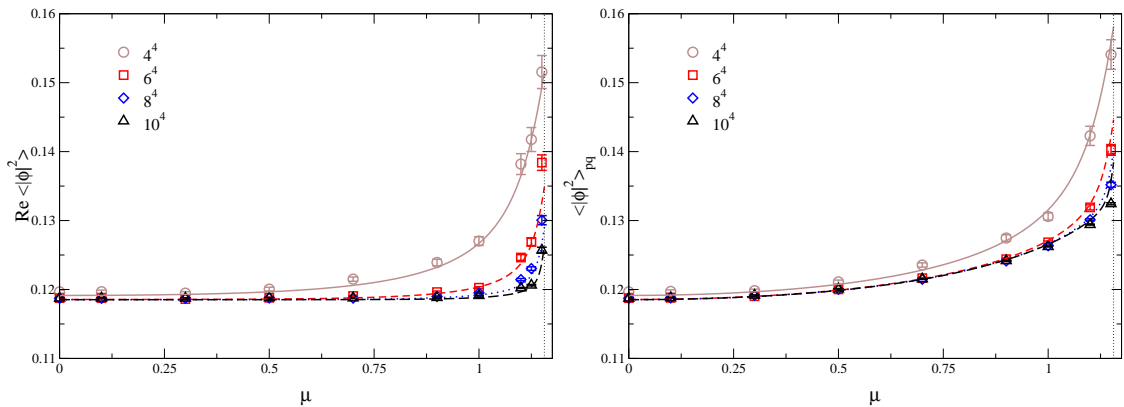
 The lines represent the mean field results for $\bra|\phi|^2\ket$ in the 
full (left) and phase quenched (right) theories for various lattice sizes, 
taking $m=\lambda=1$. The vertical dotted line indicates the mean field 
estimate for the critical chemical potential. The data points are obtained 
with Langevin simulations \cite{Aarts:2008wh}.
 }
\label{fig:ana1}
}

\FIGURE[t]{
\epsfig{figure=ana_full_dens_MF.eps,height=5.3cm}
\epsfig{figure=ana_phq_dens_MF.eps,height=5.3cm}
 \caption{
As in Fig.\ \ref{fig:ana1} for the density  $\bra n\ket$.
}
\label{fig:ana2}
}

\FIGURE[t]{
\epsfig{figure=ana_avphase_MF.eps,height=5.3cm}
\epsfig{figure=ana_Deltaf_MF.eps,height=5.3cm}
 \caption{ 
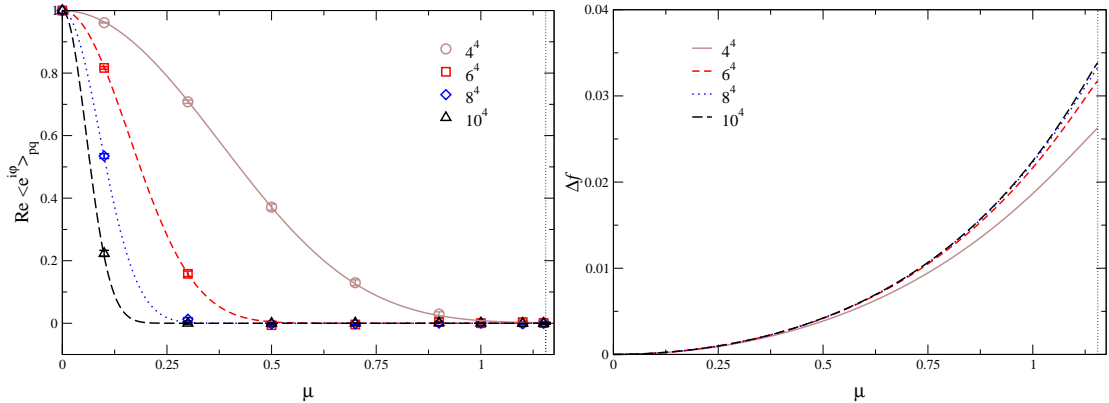
 Left: as in Fig.\ \ref{fig:ana1} for the average phase factor in the 
phase quenched theory $\bra e^{i\varphi}\ket_{\rm pq}$. Right: difference 
$\Delta f$ between the free energy densities of the full and the phase 
quenched theories in the mean field approximation.
 } 
\label{fig:ana3} 
}

We have solved the gap equation in the Silver Blaze region and used the 
outcome to compute $\bra|\phi|^2\ket$ and $\bra n\ket$ in the mean field 
approximation as a function of chemical potential for different lattice 
sizes. The results are shown in Figs.~\ref{fig:ana1} and \ref{fig:ana2} 
respectively, for $m=\lambda=1$. The vertical dotted lines indicate the 
mean field estimate of the critical chemical potential. In the full theory 
(figures on the left) the expected $\mu$ independence emerges in the 
thermodynamic limit. In the phase quenched theory (figures on the right), 
observables depend on $\mu$, since there is no Silver Blaze feature, see 
Appendix \ref{sec:exact}. Also shown in these plots are data points 
obtained from the numerical solution of the Langevin process, with 
stepsize $\eps=5\times 10^{-5}$ \cite{Aarts:2008wh}. We observe 
surprisingly good agreement between the mean field and the nonperturbative 
results for all values of the chemical potential and all lattice sizes 
considered, indicating that the mean field approximation captures the most 
relevant interactions.
 
In Fig.\ \ref{fig:ana3} we show the average phase factor in the phase 
quenched theory $\bra e^{i\varphi}\ket_{\rm pq}$ (left) and the difference 
$\Delta f$ between the free energy densities, given in Eq.\ 
(\ref{eq:deltaf}), again after the replacement $A_p\to \cA_p$. As already 
mentioned, the sign problem is severe in the thermodynamic limit: taking 
e.g.\  
$\Delta f\sim 0.02$ and a lattice volume $\Omega=10^4$, we find that the 
average phase factor $\bra e^{i\varphi}\ket_{\rm pq}=e^{-\Omega\Delta f}$ 
is indeed exponentially small.

To conclude this section, we note that it is straightforward to adapt the 
stationary solution of the extended Fokker-Planck equation, constructed in 
Sec.\ \ref{sec:fokker}, to the mean field approximation discussed here. 
Since in the mean field approximation only two-point functions appear, the 
mean field probability distribution remains of the form (\ref{eq:rho}) 
with the simple replacement $A_p\to \cA_p$ (or $m^2\to M^2$). Existence of 
the Fokker-Planck distribution in the Silver Blaze region now requires 
$\cA_p>0$.

\section{Conclusion} 
\label{sec:con} 
\setcounter{equation}{0} 

In order to further understand the applicability of complex Langevin 
dynamics for theories with a complex action due to finite chemical 
potential, we have studied the relativistic Bose gas in the Silver Blaze 
region analytically. Ignoring interactions, we have investigated 
convergence and stability, and constructed the stationary solution of the 
extended Fokker-Planck equation. We explained why this real probability 
distribution is nonlocal in real space. Subsequently, interactions were 
included on the mean field level and the fixed point of the mean field 
Langevin equations with finite stepsize was derived. We gave a comparison 
between the mean field predictions and the nonperturbative numerical data 
from Ref.\ \cite{Aarts:2008wh} in the Silver Blaze region. Surprisingly 
good agreement was found for all values of the chemical potential 
considered, including finite size effects, indicating that the mean field 
approximation captures the most important effects of the interactions. We 
have demonstrated analytically that the sign problem is severe for lattice 
volumes used in this study. From the combination of results obtained here 
and in Ref.\ \cite{Aarts:2008wh}, it can be argued that complex Langevin 
dynamics in the Silver Blaze region is well understood in this theory.

One obvious next step is to extend the analysis to the high-density phase, 
which requires the introduction of the mean field 
$\bra\phi_{a,x}^\rmR\ket$ (note that $\bra\phi_{a,x}^\rmI\ket =0$). 
Finally, it would be interesting to apply mean field approximations to 
other theories as well, especially in combination with numerical studies.
In particular, this would be useful for QCD with static quarks 
\cite{Aarts:2008rr}.

                                                                                
\acknowledgments

 I thank Kim Splittorff, Ion-Olimpiu Stamatescu and Simon Hands for their 
interest and discussion. This work is supported by an STFC Advanced 
Fellowship.


\appendix

\renewcommand{\theequation}{\Alph{section}.\arabic{equation}}

\section{Dispersion relation}
\label{sec:exact}
\setcounter{equation}{0}

 The propagator  corresponding to the action (\ref{eq:S0}) is
 \be
 G_{ab,p} = \frac{\delta_{ab} A_p + \vareps_{ab} B_p}{A_p^2+B_p^2}.
 \ee
Dispersion relations follow from the poles of the propagator, 
taking $p_4=iE_\pv$. We find
\be
 \cosh E_\pv(\mu) = \left(1+\half \hat \om_\vecp^2\right) \cosh\mu\pm 
\sqrt{ 1+\frac{1}{4} \hat \om_\vecp^2}\sinh\mu,
\ee
where 
\be
 \hat \om_\vecp^2 = m^2+  4\sum_i\sin^2\frac{p_i}{2}.
\ee
This can be written as
\be
 \cosh E_\pv(\mu) = \cosh \left[E_\pv(0) \pm \mu\right],
\ee
such that the (positive energy) solutions are 
\be
 E_\pv(\mu) = E_\pv(0) \pm \mu,
\ee
 just as in the continuum theory. Lattice discretization effects only 
appear in the dispersion relation at zero chemical potential, $E_\pv(0)$. 
The critical $\mu$ value is $\mu_c=E_\vecnul(0)=2\mbox{asinh}(m/2)$, so 
that one mode becomes exactly massless at the transition.

The phase quenched theory corresponds to $B_p=0$. In 
the formal continuum limit, the phase quenched theory is a theory with a 
real action and mass parameter $m^2-\mu^2$.  The dispersion relation is 
\be
 \cosh E_\pv(\mu) = \frac{1}{\cosh\mu}\left(1+\half \hat 
\om_\vecp^2\right),
\ee
 corresponding to $E_\pv^2(\mu) = m^2-\mu^2+\pv^2$ in the continuum limit, 
as anticipated.

These results are easily extended to the self-consistent mean field approximation, where the mass parameter receives a tadpole correction and is replaced by $M^2 = m^2 +4\lambda\left\bra|\phi|^2\right\ket$.


\end{document}